\documentclass[12pt]{article}

\def\beq{\begin{equation}}
\def\eeq{\end{equation}}
\def\barr{\begin{eqnarray}}
\def\beqa{\begin{eqnarray}}
\def\earr{\end{eqnarray}}
\def\eeqa{\end{eqnarray}}
\def\winf{W_{1+\infty}\ }
\def\u1{\widehat{U(1)}}
\def\v{V\,}
\def\w{W\,}
\def\vb{{\overline V}\,}
\def\wb{{\overline W}\,}

\newcommand{\nl}{\nonumber \\}

\begin{document}

\begin{titlepage}

\begin{center}
\hfill  \quad  \\
\vskip 0.5 cm
{\Large \bf Laminar flow of charged quantum fluids of the Calogero-Sutherland universality class }

\vspace{0.5cm}

Federico~L.~ BOTTESI$^a$ ,\ \ Guillermo~R.~ZEMBA$^{b,}$\footnote{
Fellow of Consejo Nacional de Investigaciones Cient\'{\i}ficas y T\'ecnicas,Argentina.}\\

{\em $^a$Facultad de Ingenier\'ia, Universidad de Buenos Aires,}\\
{\em  Av. Paseo Col\'on 850,(C1063ACL) Buenos Aires, Argentina}\\
\medskip
{\em $^b$Departamento de F\'{\i}sica Teórica,GIyA,Laboratorio Tandar,}\\
{\em  Comisi\'on Nacional de Energ\'{\i}a At\'omica,} \\
{\em Av.Libertador 8250,(C1429BNP) Buenos Aires, Argentina}
{\em and}\\
{\em Facultad de Ingenier\'ia y Ciencias Agrarias,  Pontificia Universidad Cat\'olica Argentina,}\\
{\em  Av. Alicia Moreau de Justo 1500,(C1107AAZ) Buenos Aires, Argentina}\\

\medskip

\end{center}
\vspace{.3cm}
\begin{abstract}
\noindent
The effective field theory of the Calogero-Sutherland model represents a universality class of 
quantum hydrodynamic fluids in one spatial dimension. It describes   
quantum compressible fluids involving both chiralities in which the chiral density field
obeys the quantum Benjamin-Ono equation.
An extension of this theory to describe a laminar flow of the Calogero-Sutherland fluids
in a rectangular geometry with small 
transverse width and the topology of a ribbon, is considered here. 
The physical picture is based on the 
edge states in the hierarchical quantum Hall effect, which may be seen 
as a collection of parallel one-dimensional quantum incompressible fluids
moving along but confined within the transverse microscopic width of the edge of the sample.
The effective theory is thus defined as the direct product of
two one-dimensional theories of the Calogero-Sutherland class 
so that one involves motion while the other is confining. 
Charge transport may be induced by coupling the system to an external 
electromagnetic field that yields a global translation of the ground state.
The effective theory describes quantum solitonic excitations along the direction of the flow and
possesses a two-dimensional electric current density which shows a Wigner semicircle law profile in 
the transverse direction, suggesting a Poiseuille-like behavior but 
without dissipative viscous effects since the velocity of 
the fluid is not a well-defined quantum field.
This simple physical picture predicts interesting phenomena with distinctive signatures that may be 
tested in real samples. 

\end{abstract}
\vskip 0.5 cm
\end{titlepage}
\pagenumbering{arabic}
\section{Introduction} 

Charge transport in narrow structures, such as nanowires, has received interest
recently due to exotic effects that may arise in quantum
hydrodynamics regimes \cite{visc,elhyd}. These effects are quite 
ubiquitous as they arise in metals, semiconductors and graphene
systems \cite{graphene} and have been observed experimentally \cite{expgra}. From the theoretical point of view, 
proposing new physical behaviors that may be of interest for this
class of systems is appealing to further exploration.

The systems we consider in this paper are described by the effective field theory ($EFT$) of the Calogero-Sutherland ($CS$) 
model in the continuous, thermodynamic limit of large number of particles. Effective field theory methods have proven to 
be successful in describing the universal behavior of low-dimensional systems significant for condensed matter and 
theoretical physics \cite{polch}. 
The physical picture that emerges for the continuous limit of the Calogero-Sutherland model
is that of a one-dimensional quantum compressible fluid that involves two chiralities. This depiction emerges in both 
the quantum field theoric and
quantum hydrodynamical formulations of the effective theories of the model \cite{aw,bz}.
We do not address in this paper the issue of finding realistic situations in which the charge transport may be 
well described by the $CS$ flow. 

The Calogero-Sutherland ($CS$) model \cite{cal,sut}
has been proposed as a compelling theoretical laboratory for 
describing quantum effects in $1D$ systems, typically embedded in
$3D$ space. The interesting features originated in the properties of the $1/r^2$ 
interaction in $3D$ space term are related to the central
potential discussion of the same dependence of \cite{lanlif}.
The $EFT$ of the $CS$ model in the thermodynamic limit has been
studied in terms of the generators of 
the $\winf$ dynamical symmetry \cite{shen,kac1,wref}, and has been formulated 
in \cite{clz,cfslz,flsz} (for alternative formulations, see also \cite{khve,amos}). 
It describes the lowest energy (``gapless'') momentum fluctuations of the Fermi surface of the system. 
Furthermore, a change in the representation of the symmetry algebra 
allows for the inclusion of non-perturbative effects due to the interaction,
in agreement with Luttinger's theorem \cite{hald1}. The advantages 
granted by the algebraic structure are the
main reason for writing the $EFT$ in terms of the $\winf$ generators.
The so obtained $EFT$ is actually an (extended) conformal cield theory ($CFT$) \cite{bpz}
which naturally incorporates the concept of the {\it bosonization} of the
lowest energy fluctuations of the Fermi surface as the relevant
semiclassical degrees of freedom \cite{boson} (see also \cite{kaya}). 
The same $EFT$ describes general Luttinger systems as well \cite{voit}.

In this paper we will consider a spatial planar geometry with the shape of a long strip, with 
charge transport along the extended dimension. 
Quantum hydrodynamic motion in this geometry is
assumed to be described by a laminar flow, in which a series
of quasi-$1D$ fluids are put together so as to gather a collective
parallel motion. This idea has, of course, a semiclassical 
picture behind. At the quantum level, the 
edge states in the hierarchical quantum Hall effect provide
an example of parallel one-dimensional quantum incompressible fluids moving within the small size of the edge of the sample.
The $EFT$ for these systems have been proposed to be direct products of $1D$ $EFT$s \cite{juerg} to describe
the parallel motion, while the transverse confining potential defines the small region of the edge of the
sample. 
On the other hand, these quantum fluids are incompressible and the number of independent ones is finite unless 
the limit of half-filled Landau is considered. 
By analogy with this example, we propose that 
the appropriate $EFT$ for the laminar flow regime would be given 
by the direct product of two $1D$ $EFT$s of the $CS$ universality class: one of them describing charge propagation along 
the extended dimension while the other one yields confinement (no motion) across the perpendicular narrow channel. 
Both theories should be subject to the fermion number conservation constraint, however.
In order to achieve this, one has to consider two different definitions of the $CS$ models corresponding
to the two orthogonal directions: the topology of the longitudinal theory is periodic (a spatial circle) while
that of the transverse one is the real line, with a harmonic confining potencial (namely, the Surtherland and 
Calogero models, respectively \cite{cal,sut}).
Charge transport may be produced in one chiral component of the $EFT$ by coupling the system to an external electromagnetic 
field.
One aim of this work is to provide an independent framework
based on $EFT$s for discussing systems with transverse narrow dimensions. The application of these ideas to concrete 
physical systems, such as nanowires at temperatures low enough and with no external magnetic fields applied so that a ballistic 
regime of charge transport is expected to be valid, is still an open issue and will not be addressed in detail here.  

This paper is organized as follows: in section 3 we present the geometry considered,
in section 4 we review the $EFT$ of the $CS$ model for the long dimension, whereas
section 5 does the same for the smaller dimension and presents the complete $EFT$. In section 6 we consider the 
the electromagnetic response of the entire system to an external fields, and in section 7 we
discuss the semiclassical limit of it. We conclude the paper in section 8 with
discussions and conclusions. 

\section{General considerations}
The contents of this paper should be considered as a continuation and generalization of the results presented in \cite{flsz}.
We propose here an extension of the $EFT$ given there as well as in \cite{aw}, which has been shown
to be precisely the same \cite{bz}. 
We first conceive the idea of a narrow $2D$ channel extending the domain of application of the 
$EFT$ of the $CS$ model. We follow the example of the edge states in the $QHE$ hierarchical $EFT$s.
For filling fractions of the form $\nu = m/\left( mp \pm 1 \right) $, with $m$ a positive integer and $p$ a positive
even integer, $EFT$s of the form of a direct product of $m$ propagating modes have been proposed to 
successfully describe them \cite{juerg}. 
These modes propagate as $c=1$ free bosonic $CFT$s in parallel or antiparallel regimes, with no friction and confined 
to the small region of the boundary of the sample (of the size of the $UV$ cutoff of the theory). This successful
picture has been entirely formulated in the second quantized formulation. Later research has shown that this
picture could be realized in first quantized formulations as well (see, {\it e.g.}, \cite{cmsz}). We stress here that
in the formulation of an $EFT$ there is no need to start from a microscopic, first quantized model and take the
corresponding thermodynamic limit so that the theory is formulated \cite{polch}. However, for illustrative 
reasons it may be convenient at times to choose such a model. For example, in the case of the $1D$ Ising 
model (see the textbook in \cite{bpz}). Having this idea in mind, we first propose a $2D$ first quantized model
that could be a representative of the $EFT$ presented at the end of Section 5, which, in our logic of thinking 
it is the starting point. That is, we conceive the $EFT$ inspired in the $QHE$ hierarchical ones, and then work
backwards our logic to provide a consistent picture based on a first quantized model. We have chosen to present 
our proposal in this way for aiming at a larger community of interested readers, who perhaps are not familiar with the
ideas and methods of $EFT$s. This approach may have, however, some criticism related to the assumptions made
for presenting the first quantized model. We emphasize here that we just provide this model to be a representative
of the universality class that describes the $EFT$, constructed in analogy to well known consistent ones.  
\section{The geometry of 2D thin channels}
In this section, we provide a basic first quantized setting for extending the results of the $1D$ $EFT$ 
of the $CS$ model presented in \cite{flsz,aw}. It should be considered within the scope and imitations discussed 
in the previous section. 

We consider a planar rectangular space geometry, of length $L_x$ 
along the $x$ direction and $L_y$ along the $y$ direction. Here
$x$ and $y$ are the two Euclidian coordinates of the plane. 
This geometry is meant to describe a flow of charges along the 
$x$ direction and a confined density along the $y$ direction. 
We consider the system to be periodic in the $x$
direction and with no periodicity in the $y$ direction. 
Topologically, this region is, therefore, a ribbon.
In terms of the $QHE$ analogy discussed above, the $x$ direction is analogous to the propagation dimension
of the edge states and the $y$ direction, the small region of the edge size to which the dynamics is confined. 

Spinless charged particles, the fermionic degrees of freedom of the $CS$ model, are the dynamical objects assumed to 
move in this spatial domain. It is given by a matrix array of $N_x$ files and $N_y$ columns, with $N_x \gg N_y$. 
Each of the finite size matrix elements is seen as a dynamical degree of freedom.
The total number of particles $N$ and area $A$ are thus given by $N = N_x N_y$ and $A = L_x L_y$, respectively.
That is, consistently with our picture, only global quantities relate both dimensions. Intuitively, in the resulting
$EFT$, we expect a very large number of horizontal currents that are basically independent (as in the edge states).
We shall describe their dynamics in the
thermodynamic limit by means of a $(2+1)$-dimensional $EFT$,
which is constructed as the direct product of two (spatial) $1D$
ones (again, based on the $EFT$ previously developed). We therefore assume that the degrees of freedom that describe
the dynamics in the $x$ and $y$ directions are independent.
The thermodynamic limit is achieved by taking the limits $L_x \to \infty$, 
$N_x \to \infty$ such that the average number density $n_0^{(x)}=N_x /L_x$ is finite,
and, similarly, 
$L_y \to \infty$, $N_y \to \infty$ such that $n_0^{(y)}=N_y /L_y$ is finite.
One expects $ n_0^{(x)} = n_0^{(y)} $ but for the moment we keep these 
two quantities as independent for the clarity of the discussion.
The rectangular domain is therefore decomposed as a matrix (mosaic) of $N$ tiles
of sizes $\Delta x = 1/n_0^{(x)}$ and $\Delta y = 1/n_0^{(y)}$,
which are the natural $UV$ cutoffs of the $EFT$.
Densities may become local quantities as well: $n_0^{(x)}\left( x \right) =dN_x /dL_x$ 
and $n_0^{(y)}\left( y \right) =dN_y /dL_y$.
The $2D$ particle number density is given by:
\beq
n^{(xy)} \left( x, y \right )\ =\ n^{(x)} \left( x \right )  n^{(y)} \left( y \right )\  .
\label{density-2d}
\eeq
Particle number densities may be turned into charge densities by multiplication by  
the charge of single particles, that we assume here to be $e$ (and that may be replaced by 
an effective value of the charge as well).
\section{The 1D EFT of the longitudinal dimension}
We first consider the $EFT$ for the $x$ (periodic) direction. 
For the sake of simplifying the notation, in this discussion 
we will drop the suffix $x$ in all relevant variables, that should be understood.  
We review here the main characteristics of the Calogero-Sutherland model \cite{cal,sut}.
Consider a system of $N$ non-relativistic 
$(1+1)$-dimensional spinless interacting fermions on a circle of length $L$, with 
Hamiltonian \cite{sut} (in units where $\hbar =1$ and $2m=1$ , with $m$ being 
the mass of the particles)
\beq
h_{CS}=\sum_{j=1}^N\ \left( \frac{1}{i} \frac{\partial}{\partial x_j}\
\right)^2\ +\ g\ \frac{\pi^2}{L^2}\ \sum_{i<j}\ \frac{1}{\sin^2
(\pi(x_i-x_j)/L) }\ ,
\label{ham}
\eeq
where $x_i$ ($i=1,\dots,N$) is the coordinate of the $i$-th particle, and $g$ is 
the dimensionless coupling constant. Ground state stability demands $g \geq -1/2$,
with both attractive ($-1/2 \leq g < 0$) and repulsive ($0 < g$) regimes. A usual 
reparametrization of the coupling constant is given by $g=2 \xi ( \xi -1)$, 
so that $\xi \geq 0$ and $0 \leq \xi < 1$ is the attractive regime and $1  < \xi $ 
the repulsive one. We shall be concerned with the repulsive regime for the rest 
of the paper. 
Fermions in the $CS$ model are usually taken as electrically neutral, but 
may be given an electric charge as well. In that
case, one has to assume that Coulomb interactions among the
fermions are screened, so that only electric coupling to an 
external electromagnetic field is of relevance \cite{zirnhal}. 
For charge balancing, a rigid neutralizing passive background may be added to the system.  

The $EFT$ of (\ref{ham}) has been obtained in 
\cite{clz}\cite{cfslz}\ by reformulating the system dynamics in terms of variables 
(fields) that directly describe that of the $1D$ Fermi surface in the thermodynamic 
limit. 
This method amounts to defining initially suitable non-relativistic fermionic fields, 
taking then the thermodynamic limit $N \to \infty$ properly on the fields and hamiltonian 
to obtain an $EFT$ that involves two sets of independent relativistic (in the sense of 
a linear dispersion relation) fermion fields that describe the low-energy fluctuations 
around each of the two Fermi points of the $1D$ effective theory. The following 
step consists in writing down the $EFT$ in terms of fields that display the $\winf$ 
symmetry. This step is crucial in our approach, as it allows to obtain the Hilbert space of 
the $EFT$ by exploiting the algebraic 
properties of the $\winf$ algebra and diagonalize the hamiltonian.
Moreover, the $\winf$ structure obtained in the original fermionic basis may be realized as well by
bosonic operators that take into account interacting fermions, as anticipated.
This procedure has been recently reviewed in \cite{bz}. For completeness we
quote here some results. 
The spatial density $n(x,t)=n_0=N/L$ is uniform and stationary, and the 
effective hamiltonian is the following operator: 
\barr
{\cal H}_{CS} &=&\left(
2\pi n_0  \sqrt{\xi}\right)^2
\left\{\left[\frac{\sqrt{\xi}}{4}\,\w_0^0
+\frac{1}{N}\,\w_0^1+
\frac{1}{N^2}\left(\frac{1}{\sqrt{\xi}}\,\w_0^2
-\frac{\sqrt{\xi}}{12}\,\w_0^0 \right.\right.
\right.\nl
&&-\ \left.\left.\left.
\frac{g}{2\xi^2}\,\sum_{\ell=1}^\infty
\,\ell~\w_{-\ell}^0\,\w_\ell^0\right)
\right]+\left(\,W~\leftrightarrow~{\overline W}\,\right)
\right\}~~~,
\label{hcsf}
\earr
where $\xi=\left(1+\sqrt{1+2g}\right)/2$ is the standard reparametrization of the coupling 
constant parameter defined after (\ref{ham}).
Note that $\xi = 1$ corresponds to the free fermion case.
Nevertheless, the relationship between $\xi $ and $g$ is not relevant when discussing 
the properties of the $EFT$, which leaves behind all the small-scale features 
of the underlying dynamics. Therefore, $\xi$ is taken as a real positive {\it independent} parameter.

The operators $~\w_{\ell}^m$
in (\ref{hcsf}) are the lowest (in $m=i+1$, where $i$ is the conformal spin \cite{bpz}) 
generators of the infinite dimensional algebra known
as $\winf$ \cite{shen,kac1}. The terms in the $~\w_{\ell}^m$ ($\wb^i_\ell$) operators describe the 
dynamics at the right $(R)$ (left $(L)$) Fermi point, respectively. 
We remark that the {\it complete factorization} of (\ref{hcsf}) into chiral
and antichiral sectors is possible for the $CS$ $EFT$ only after performing a Bogoliubov
transformation that decouples both sectors, that are generically mixed 
by backward scattering terms in the first fermionic form of the Hamiltonian obtained 
in the thermodynamic limit \cite{cfslz,flsz}.

The general form of the $\winf$ algebra is:
\beq
\left[\ \w^i_\ell, \w^j_m\ \right] = (j\ell-im) \w^{i+j-1}_{\ell+m}
+q(i,j,\ell,m)\w^{i+j-3}_{\ell+m}
+\cdots +\delta^{ij}\delta_{\ell+m,0}\ c\ d(i,\ell) \ ,
\label{walg}
\eeq
where the structure constants $q(i,j,\ell,m)$ and $d(i,\ell)$ 
are polynomial in their arguments, $c$ is the central charge, 
and the dots denote a {\it finite} number of terms involving the operators 
$\w^{i+j-2k}_{\ell+m}\ $.
The ground state $|\ \Omega\ \rangle$ is a highest-weight state 
with respect to the $\winf $ operators, namely
$\w^i_\ell |\ \Omega\ \rangle = 0$, $\ell > 0, i \geq 0$,
that is tantamount to incompressibility in momentum space. 
We remark here that the basis $\w^i_\ell$ of $\winf$ operators in 
the Hamiltonian (\ref{hcsf}) is not the original fermionic one, inherited from the 
$CS$ model, but rather a bosonic one, as we shall explain below. 
One major advantage for choosing the basis of the $\winf \times 
{\overline \winf}$ operators is that, once the algebraic content 
of the theory has been established in the free fermionic picture,
the bosonic realization (in terms of bosonic field) of the algebra can be used,
and the free value of the compactification radius of the boson
can be chosen so as to diagonalize the Hamiltonian.
This method is consequently termed as {\it algebraic bosonization} \cite{flsz}.
For the case of the $CS$ $EFT$ $c=1$, and all the relevant commutation relations are:
\barr
\left[\ \w^0_\ell,\w^0_m\ \right] & = &  c\ \xi \ell\ \delta_{\ell+m,0} ~~~,\nl
\left[\ \w^1_\ell, \w^0_m\ \right] & = & -m\ \w^0_{\ell+m} ~~~,\nl
\left[\ \w^1_\ell, \w^1_m\ \right] & = & (\ell-m)\w^1_{\ell+m} + 
\frac{c}{12}\ell(\ell^2-1) \delta_{\ell+m,0}~~~,\nl
\left[\ \w^2_\ell, \w^0_m\ \right] &=& -2m\ \w^1_{\ell+m}~~~,
\label{walg1}\\
\left[\ \w^2_\ell, \w^1_m\ \right] &=& (\ell-2m)\ \w^2_{\ell+m} -
   \frac{1}{6}\left(m^3-m\right) \w^0_{\ell+m}~~~,\nl
\left[\ \w^2_n, \w^2_m\ \right] &=& (2n-2m)\ \w^3_{n+m}
     +{n-m\over 15}\left( 2n^2 +2m^2 -nm-8 \right) \w^1_{n+m}\nonumber\\
     &&\quad +\ c\ {n(n^2-1)(n^2-4)\over 180}\ \delta_{n+m,0}~~~.\nonumber
\earr
The first and third equations in (\ref{walg1}) show
that the generators $\w^0_\ell$ satisfy the 
abelian Kac-Moody algebra $\u1$, and the generators $\w^1_\ell$ 
satisfy the Virasoro algebra, respectively.
The operators $\wb^i_\ell$ satisfy the same algebra (\ref{walg1}) 
with central charge ${\overline c}=1$ 
and commute with the all the operators $\w^i_\ell$. The complete $EFT$ 
of the $CS$ model is a $(c,{\overline c})=(1,1)$ $CFT$, but since both chiral ($R$) and
antichiral ($L$) sectors are isomorphic, we will often consider one of them for
the sake of simplicity.

The $\winf$ algebra may be realized in terms of a chiral bosonic field 
by a generalized Sugawara construction \cite{kac1} .
In fact,defining the right and left moving modes,
$\alpha_\ell$ and ${\overline \alpha}_\ell$, of a free compactified 
boson ($[\alpha_n,\alpha_m ]=\xi n \delta_{n+m,0}$ and similarly for 
the ${\overline \alpha}_\ell$ 
operators), the commutation relations (\ref{walg1}) 
are satisfied by defining $\w^i_\ell$ (we only write the expressions
for $i=0,1,2$) as:
\barr
\w^0_\ell &=& \alpha_\ell ~~~,
\nonumber\\
\w^1_\ell &=& {\frac{1}{2}} \sum_{r= -\infty}^{\infty}
:\, \alpha_{r}\,\alpha_{\ell-r}\,
:~~~,\label{mod2}\\
\w^2_\ell &=& {\frac{1}{3}} \sum_{r, s = -\infty}^{\infty}
:\, \alpha_{r}\,\alpha_s\, \alpha_{\ell-r-s}\,:~~~,
\nonumber
\earr
and analogously for the operators $\wb^i_\ell$ in terms of ${\overline \alpha}_\ell$.
The naive generalization of (\ref{mod2}) to higher 
values of conformal spin is incorrect (for example, see \cite{flsz}).

We now outline the spectrum of (\ref{hcsf}). In terms of the bosonized operators 
basis $\w^i_\ell$ the highest weight vectors,
$|\Delta N ; \Delta D \rangle_W$, 
are obtained by adding $\Delta N$ particles to the 
ground state $|\Omega \rangle_W$, and by moving $\Delta D$ 
particles from the left to the right Fermi point, {\it i.e.},
\barr
\w_0^0 ~|\Delta N ; \Delta D \rangle_W  &=&
\left(\sqrt{\xi}\,\frac{\Delta N}{2}+
\frac{\Delta D}{\sqrt{\xi}}\right)
|\Delta N ; \Delta D \rangle_W \nl
\wb_0^0 ~|\Delta N ; \Delta D \rangle_W  &=&
\left(\sqrt{\xi}\,\frac{\Delta N}{2}-
\frac{\Delta D}{\sqrt{\xi}}\right)
|\Delta N ; \Delta D \rangle_W~~~.
\label{deltandeltad}
\earr
The highest weight states $|\Delta N , \Delta D \rangle_W$
together with their descendants, denoted by
$|\Delta N , \Delta D ; \{k_i\},\{{\overline k}_j\} \rangle_W$,
with $k_1 \ge k_2 \ge \dots \ge k_r > 0$, and
${\overline k}_1 \ge {\overline k}_2 \ge \dots 
\ge {\overline k}_s > 0$.

The exact energies of these excitations in this basis are given by:
\barr
{\cal E}&=&\left(
2\pi n_0  \sqrt{\xi}\right)^2
\left\{\left[\frac{\sqrt{\xi}}{4}\,Q+\frac{1}{N}
\left(\frac{1}{2}\,Q^2+k\right)
+\frac{1}{N^2}\left(\frac{1}{3\sqrt{\xi}}\,Q^3
-\frac{\sqrt{\xi}}{12}\,Q\right.\right.\right.
\label{eba} \\
&&+\ \left.\left.
\frac{2k}{\sqrt{\xi}}\,Q + \frac{\sum_j k_j^2}{\xi}
-\sum_j \left(2j-1\right) k_j\right)\right]
+\left(Q\ \leftrightarrow \ {\overline Q}~,
{}~\{k_j\} \ \leftrightarrow \ \{{\overline k}_j\} \right)\Bigg\}~~~,
\nonumber
\label{exacte}
\earr
where
$$ k \ =\ \sum_j k_j ~~~~,~~~~
{\overline k} \ =\ \sum_j {\overline k}_j
$$
and the eigenvalues of $\w_0^0$ and $\wb_0^0$ are, respectively:
\beq
Q=\sqrt{\xi}\,\frac{\Delta N}{2}+
\frac{\Delta D}{\sqrt{\xi}}~~~~,~~~~
{\overline Q}=\sqrt{\xi}\,\frac{\Delta N}{2}-
\frac{\Delta D}{\sqrt{\xi}}~~~.
\label{Q}
\eeq
Moreover, the integer numbers $k_j$ are ordered
according to $k_1\geq k_2\geq \dots \geq 0$
and are different from zero only if $j << \sqrt{N}$ (and analogously 
for the ${\overline k}_j$). We remark here that these are the (renormalized) 
excitation energies in the $EFT$, which differ from those obtained
in first quantized studies by an infinite subtraction \cite{flsz}.   

The Hilbert space of the theory can be recognized as that of a
$(c,{\overline c})=(1,1)$ $CFT$s with extended 
symmetry $\winf \times {\overline \winf}$ and chiral ($R$) and
antichiral ($L$) sectors that are isomorphic \cite{clz,cfslz,flsz} . 
It coincides with that of the free non-chiral
bosonic field (which is the sum of chiral and antichiral bosons) 
compactified on a circle of radius $r = \frac{1}{\sqrt{\xi}}$,
as noted in \cite{kaya}. 
The partition function of this $CFT$ is known to be invariant 
under the duality symmetry $r \leftrightarrow 1/(2r)$, which is
equivalent in the $CS$ $EFT$ to $\sqrt{\xi} \leftrightarrow 2/\sqrt{\xi}$.
The action of this mapping on the charges (\ref{Q}) is to interchange 
$\Delta N \leftrightarrow \Delta D$, such that $Q \leftrightarrow Q $ ,
 ${\overline Q} \leftrightarrow -{\overline Q}$.
See \cite{bz} for more details.


Therefore the $EFT$ describes a uniform density ground state that may have 
solitons (located lumps or valleys) and fluctuations on top of them as the complete set of 
low-lying excitations.
The existence of soliton-like excitations in the direction of the flow
is one of the qualitative features we remark in this work, and 
have origin in the compressible nature of the fluid. 
Unusually for a $CFT$, we have higher order corrections in $Q$ beyond $Q^2$ in the Hamiltonian,
because the universal finite-size corrections are of order $1/N$ (see also \cite{soti}).
\section{The 1D EFT of the transverse dimension}
We now consider the $EFT$ for the $y$ direction, with the strip located
symmetrically around the origin. A similar analysis can be performed if the
fermions are located on the real line, rather than on a circle.
Again we will drop the suffix $y$ in all relevant variables.
In this case, the Hamiltonian is \cite{cal}\cite{sut}
\beq
{\hat h}_{CS}=\sum_{j=1}^N\ \left[\left( \frac{1}{i}
\frac{\partial}{\partial y_j}\ \right)^2\
+\ \omega^2\ y^2_j\ \right]\ +\
g\ \sum_{j<k}\ \frac{1}{(y_j -y_k)^2}\ .
\label{hamc}
\eeq
Here $\omega$ is the strength of a confining harmonic potential,
and $g$ the dimensionless coupling constant ($g \geq -1/2$).
We remind the construction of the effective theory done for (\ref{ham}).
Consider first the free case ($g=0$). The single-particle wave functions and spectrum 
are now given by
\barr
{\hat\varphi}_n(y)\ & = &\ {\cal N}_n\ H_n(\sqrt{\omega} y)\
\exp\left(- \frac{\omega y^2}{2}\right)\ ,
\quad {\cal N}_n = \left( \frac{\omega}{\pi} \right)^{1/4}
\frac{1}{2^{n/2} \sqrt{n!}}\ ,\label{showf} \\
{\hat\epsilon}_n\ & = &\ 2\omega\ \left(\ n\ +\ \frac{1}{2} \right)\ ,
\qquad n=0,1,2, \dots\ ,
\label{hermi}
\earr
where $H_n(y)$ are the Hermite polynomials. For convenience, we also define 
${\hat\varphi}_n(y,t) \equiv {\hat\varphi}_n(y) \exp(-i{\hat\epsilon}_n t)$. 
Next, we introduce a second quantized field
\beq
{\hat \Psi} (y,t)\ =\ \sum_{n=0}^{\infty} c_n\ {\hat\varphi}_n (y,t)\ ,
\qquad \{\ c_k , c^{\dag}_l\ \}\ =\ \delta_{k,l}\ .
\label{fielc}
\eeq
The ground state is
\beq
|\ \Omega\ \rangle\ =\ c^{\dag}_{N-1} c^{\dag}_{N-2} \dots
c^{\dag}_1 c^{\dag}_0\ |\ 0\ \rangle\ .
\label{gscal}
\eeq
The ground state numerical density $n(y)$
in the thermodynamic limit satisfies the {\it semicircle law}
(see, {\it e.g.},\cite{sut}):
\beq
n(y)\ =\ \frac{\omega}{\pi}\ \sqrt{y^2_0\ -\ y^2\ }\ ,\qquad
y^2_0\ =\ \frac{2N}{\omega}\ ,
\label{semi}
\eeq
for values $y^2 \le y^2_0$ and zero otherwise. The semiclassical
location of the spatial boundary of the ground state is given by
$\pm y_0$, and the (semiclassical) value of the Fermi momentum is
$p_F=\sqrt{2N\omega}$.
The thermodynamic limit of $N$ large and $\omega$ small,
is taken such that $p_F$ and the value of the density
at the origin $n(0)=\sqrt{2N\omega}\ / \pi$ are finite quantities.
Note that the specific analytic form of (\ref{semi}) depends upon the 
specific (harmonic) form of the confining potential in (\ref{hamc}).

Imposing now the condition $y_0=L_y/2$ given by the considered geometry,
we may express (\ref{semi}) as:
\beq
n(y)\ =\ \frac{4n_0^{(y)}}{\pi}\ \sqrt{1 -\ 4\frac{y^2}{L_y^2}\ }\ ,
\label{semi2}
\eeq
with $\omega = 8n_0^{(y)}/L_y$. Note that in the thermodynamic limit $\omega \to 0$.
This is the frozen oscillator limit, in which the size of the ground
state wave function becomes large and no $CS$ particles move along the $y$
direction. When it exceeds the size of the system $2L_y$
no charge transport is expected, even in the 
presence of an external electromagnetic field. 
This limit corresponds 
to the weak magnetic field regime of the Quantum Hall Effect ($QHE$), 
for which there is no Hall current \cite{qhewf}. 
Therefore, the $EFT$ along the $y$ direction describes quantum charge confinement.
In physical systems, this may be achieved by Coulomb blockade or similar
mechanisms. 

The spectrum of this $EFT$ has been studied in \cite{clz} up to and including 
order $1/N$ and shown to coincide with the corresponding one in (\ref{exacte}) 
provided that $(2\pi n_0)^2$ is replaced by $2\omega N$. This has been verified
to first order in $g$ in $\xi$. 
It is clear that in our geometry this level of accuracy is
sufficient since $N_x \gg N_y$: this means that the spectrum of low-lying energies
of the entire system 
will be dominated by those corresponding to the $EFT$ in the $x$ direction, as 
implied by the form of (\ref{exacte}). Moreover, we would not couple the system to external
electromagnetic fields that could induce a translation along the $y$ direction.

We are now in a position to discuss the $2D$ density (\ref{density-2d}) of the complete
$(2+1)$-dimensional $EFT$. The combined density is given by:
\beq
n^{(xy)} \left( y \right )\ =\ 
 \frac{4 n_0^{(x)} n_0^{(y)}}{\pi }\ \sqrt{1 -\ 4\frac{y^2}{L_y^2}\ }\ . 
\label{density-2d2}
\eeq
This is the second of the highlights of this work: the consistent density of the
(charged) fluid in a static (non moving) condition is given by this 
half-cylinder shape, displaying a vanishing density at the narrow edge.
Particle number density (\ref{density-2d2}) is turned into charge density by 
multiplication by the factor $e$. This can be verified by noting that the total 
charge of the system is $Ne$, and integration of (\ref{density-2d2}) over the
sample yields $N$.
\section{Quantum charge transport mechanism}
So far we have considered the density of charge 
of the system in equilibrium. We now focus on the transport of charge 
mechanism so that electric currents in the quantum
domain and in the ballistic regime may be computed.
We assume that the particles of the $CS$ model are charged, such that the
particle become charge densities,
after multiplication by a factor of $e$. 
At first sight we could consider applying to the system an external electric field along 
the $x$ direction and produce a current response 
as a global bulk motion. We will postpone to the next section considering this 
view, as it is a semiclassical perspective and we are firstly interested in 
determining the quantization of the electric currents,
as expected in nanowire-type systems. 

Charge transport may alternatively be obtained as a response 
to an induced electric field, that breaks the parity symmetry along the $x$ axis,
by coupling the system to an external {\it magnetic} field. 
This mechanism has been discussed in the literature of the $QHE$ 
(the so-called `Laughlin thought experiment´ or the Abelian charge anomaly
in a fermion system) \cite{laugh,cdtz}, and has the virtue
of being intrinsically quantum mechanical, given the quantization of the magnetic flux. 
Here we would like to present a variation of this mechanism
in which the charge transport takes place in momentum space 
between the two Fermi points, but implies an analogous one in space.
Note, however, that in this picture magnetic fields are very small,
and appear as a part of the thought experiment. There is no external strong
magnetic field applied to the system as in the case of the $QHE$.
In this picture, electric currents are not due to global motion 
of the bulk charge density but rather to solitons moving on top of the 
static density profile (see \cite{aw} also). Unlike the case of the
$QHE$, the two induced currents do not add up to zero but rather sum up
because of the different sings in the electric charges of the corresponding 
pairs of electric currents
(this comparison can be made by considering the limiting case of an annulus geometry 
that collapses into a ring \cite{cdtz} ).

\noindent
{\it i) The 1D case:} for simplicity consider first free fermions ($\xi = 1$). 
In $1D$ momentum space, the thought experiment is based on 
the creation of 
a pair of excitations $(Q,{\overline Q})=(1,-1)$, which is interpreted 
as a rigid translation of the Fermi sea (segment) one unit of $CS$ linear momentum 
$ \Delta p_x = \left( 2 \pi /L_x \right) $ directed from $L$ to $R$. 
One naturally expects that this momentum transfer manifests itself in
space as a motion of electric charge $e\Delta Q = e\left( Q -{\overline Q}\ \right) /2$ 
giving rise to an electric current.
This picture may be confirmed and expanded by considering its alternative
description in coordinate space, which we take to be a strip of width $\Delta y$ and
length $L_x$. 
We first define the soliton ($S$) or antisolition ($AS$) as 
a spike or hole in the number density of height $n_0$ and width (size) $\Delta x $,
respectively. From the previous discussion, their electric charges are 
$e$ and $-e$, respectively.
An external electric field along $x$ induced from a varying magnetic flux
creates a pair $S-AS$ $(Q,{\overline Q})=(1,-1)$
at nearby points $(x, \Delta y /2)$ and $(x, -\Delta y /2)$, respectively. 
The energy for creating this pair
is provided by the electric field,
as in the creation process of an electron-positron
pair out of a photon in $QED$ (we shall expand on this later). 
This process preserves particle number ($\Delta N = 0$) 
but induces a kinematical linear momentum $\Delta p_x$ in the system, which
may be calculated from the canonical momentum conservation law:
\beq
\left[ m v_x + e A_x(\Delta y /2) \right] + 
\left[ -m {\overline v}_x - e A_x(- \Delta y /2) \right]=0\ ,
\label{pcons}
\eeq
where $v_x$ and $-{\overline v}_x $ are the (positive) velocities of 
the $S-AS$ pair (respectively),
$A_x (y)$ is the magnetic vector potential corresponding to a uniform magnetic field
in the $-z$ direction, $B_0 =dA_x/ dy$ (choosing a gauge that is natural for this geometry) 
and $m$ the mass of a $S$ or $AS$,
which in our units is $1/2$ (same as the mass of the $CS$ particles, because the size
of the $S$ or $AS$ is $\Delta  x$ such that the number of $CS$ particles
in this length is one). Therefore, $ \Delta p_x = \Delta v_x / 2 = e \Delta y B_0$,
with $ \Delta v_x = v_x - {\overline v}_x $. Considering the magnetic field
to be minimal, that is $ \Phi_0 = B_0\  L_x\   \Delta y \ $ ,
 with $\Phi_0 = \pi /e$ the quantum unit of magnetic flux, we obtain 
$\Delta v_x =  \left( \pi /L_x \right) $, or $\Delta p_x =  \left( 2 \pi /L_x \right) $,
as expected.
The momentum transferred to the system produces a transport of charge along $x$. 
Once the $S-AS$ pair is created, each one of its constituents moves under the action 
of the external electric field giving rise to two counter-propagating modes 
that sum up as two identical electric currents flowing from $L$ to $R$ (note that
the actual physical current goes in the opposite direction, due to Lenz's law).
Therefore, the role of the electric field is two fold: it creates the $S-AS$
pair and then an electric current out of it. 

Returning to the threshold energy necessary for the $S-AS$ pair creation,
it may be calculated from the condition $2m c^2 = q \Delta V$,
where $c$ is the speed of light and $\Delta V$ is the potential difference 
along the length $L_x$,
which may be obtained from Faraday's law as 
$\Delta V = \Delta \Phi / \Delta t=  \Phi_0 v/ L_x = 2 \pi^2 n_0^{(x)} 
/(e L_x)$. Writing $v = \alpha c $, that is, the Fermi velocity as a fraction $\alpha$ 
the speed of light, we get the threshold condition $\alpha^2 = 1/N_x$. 
In realistic systems, $\alpha \simeq 10^{-2}$ impliying $N_x = 10^4$ as a minimal
requirement for creation of a $S-AS$ pair.

\noindent
{\it ii) The 2D case:} Consider now the $2D$ system decomposed in infinitesimal 
$1D$ strips along the $x$ direction, with transverse size and number of degrees 
of freedom $dy$ and $dN_y$, respectively. 
We follow and extend the thought experiment of the $1D$ case, considering now the global 
system as a circular ribbon embedded in $3D$ space,
where the $x$ coordinate of the previous discussion corresponds to the angular 
variable $R \theta$ ($R$ being the radius of the
circle) and the $y$ coordinate to $z$ in the new cylindrical coordinate system, 
along which the magnetic field is located.
Considering the adiabatic process of adding one unit of quantum 
flux $\Phi_0 $, then Faraday's law implies a charge transport along each 
infinitesimal ring in which the strip is decomposed. 
One obtains (in absolute value)
\beq
\Delta \Phi =  \Delta Q\  dq/dG\ , 
\label{faraday}
\eeq
where $\Phi$ is the 
magnetic flux subtended by the ring,
$\Delta Q = \left( Q -{\overline Q}\ \right) /2$ is the soliton number 
transported from $L$ to $R$, $dq$ is the electric charge and $dG$ the 
differential electrical conductivity, expected,
by dimensional reasons, to be
a multiple $\nu $ of the quantum of electric conductance 
$G_0 = e^2/(2 \pi) $, {\it i.e.}, $G= \nu G_0$.
Consider an element of area $dx dy $ in the strip.
Note that $dq = \left( e / \xi \right)\ n^{(xy)} \left( y \right )\ dx dy = 
\left( e / \xi \right)\ dN_x dN_y$ 
and that, by using the standard expression for electric 
conductance of a $2D$ conductor 
$dG = \nu G_0\ n^{(xy)} \left( y \right )\ dy/L_x = \nu G_0\ dN_x dN_y$. 
Therefore, from (\ref{faraday}), we see that one unit of quantum flux,
$\Delta \Phi = \Phi_0 $, yields a transport of one unit of charge $\left( e / \xi \right)$ 
from $L$ to $R$, $\Delta Q = 1$, for 
\beq
G =  e^2 / \left(\pi {\xi} \right) =2 G_0\ /{\xi} , \qquad \nu = 2/{\xi}, 
\label{qcond}
\eeq
in agreement with the previous qualitative picture of two identical currents.
Fermi statistics impose $\xi = m = 1,3,5,\dots$ \cite{juerg}.

The charge transport process just described has an interpretation in terms of 
the spectrum of the Hamiltonian (\ref{hcsf}), 
which, without loss of generality, we present in the fermionic picture ($\xi = 1$): 
$\Delta D$ in (\ref{deltandeltad}) corresponds exactly to the same charge transport 
from $L$ to $R$ considered in the thought experiment, {\it i.e.}, $\Delta D = \Delta Q$. 
A consequence of this equivalence is that {\it the energy and linear 
momentum transfer from the electromagnetic field to the system is
independent of the specific process considered}. 
Therefore, a minimal rigid translation of the
bulk along the $x$ axis may be obtained as well by setting 
$\Delta D = 1$ and $\Delta N = 0$ to keep the charge (particle number) 
unmodified.  
The energy of this translation is obtained from (\ref{exacte}),
noting that it corresponds to $Q =1$, $ {\overline Q}=-1$ and $k = {\overline k} = 0$,
yielding (after restoring explicitly $N_x$ in place of $N$)
\beq
d{\cal E}=\left(2\pi n_0^{(x)} \right)^2 dN_y/N_x\ .
\label{tarnse}
\eeq
For the rest of the discussion we 
will consider the effects of the minimal velocity induced by one magnetic flux quantum.
Nevertheless, within the limits of the validity of the $EFT$, the maximum number of magnetic 
quanta allowed is of the order of $\sqrt{N_x}$ and this maximal case could be also considered
as well. Therefore, the minimal  energy associated to a global charge transport along 
the entire ribbon is:
\beq
{\cal E} = \left(2\pi n_0^{(x)} \right)^2 N_y/N_x\ .
\label{tarnsefi}
\eeq
Had we consider the interacting case with $\xi \neq 1$, we would have obtained exactly 
the same results because the extra factor of $\sqrt{\xi}$ in $Q$ (\ref{Q}) cancels 
out with the corresponding one in the $1/N$ term of the energy (\ref{exacte}).
\section{Semiclassical charge transport mechanism}
We now consider the regime of semiclassical electric conduction
(for simplicity we consider here $\xi = 1$, but the general case yields
similar results after multiplication of the current density by the
factor $1/{\xi}$) .
We shall assume that the system may support global charge transport by 
translation of the charge density along the $x$ axis as a response to
an external electric field applied. 
The semiclassical $2D$ electric current density in the $x$ direction,
which is given by 
\beq
K_x \left( y \right)\ =\ e\ n^{(xy)}\left( y \right)\ V_x\ ,
\label{currdens}
\eeq
where $V_x$ is the global velocity of the charged fluid, 
which can be calculated by equating the energy ${\cal E}$ transferred from
the electromagnetic field to the kinetic energy of the entire system,
of mass $M = N m$, {\it i.e.}, $ {\cal E} = NV_x^2/4$ (the unit of mass is $1/2$), 
yielding $ V_x = 2 v / N_x$, with the Fermi
velocity given by $v = 2 \pi n_0^{(x)}$. 
After replacing $n_0^{(x)}  n_0^{(y)}= n_0^{2D}= N/A$, the uniform $2D$ 
particle number density, we obtain 
\beq
K_x \left( y \right)\ =\  
\frac{16  e n_0^{2D} }{L_x}\ 
\sqrt{1 -\ 4\frac{y^2}{L_y^2}\ }\ .
\label{currdensfin}
\eeq
This expresses the Poiseuille-like behavior of the system \cite{visc},
{\it i.e.}, a non-dissipative quantum viscous effect transverse to the direction of the flow
not unlike the $QHE$ viscosity \cite{qhevis} (see also \cite{ga}). The Poiseuille flow refers 
to viscous classical fluids with dissipation and exhibits a velocity profile of the 
form $v(u)=v_0(1-u^2)$,
with $u$ a dimensionless transverse coordinate. Here we got a law for the current density 
of the form $K_x(u)=K_0\sqrt{1-u^2}$ instead, with no dissipation implied. We remark
that the velocity is not a natural variable in the $EFT$ considered here (see \cite{aw} for
an alternative). Nevertheless, in both cases the corresponding quantity, $v(u)$ or $K_x(u)$ is 
an even function of $u$ that vanishes at $u=\pm 1$. 

The current density (\ref{currdensfin}) may be integrated along the $y$ direction 
so as to obtain the electric current intensity along the $x$ axis:
\beq
I_x  = \ 4\pi\ e\ n_0^{2D}\ \frac{ L_y }{L_x}\  .
\label{incurr}
\eeq
After restoring the physical units to this expression by multiplication
by the factor $\hbar / (2m)$, we obtain our final expression for $I_x$:
\beq
I_x  = \ e\ n_0^{2D}\ \left( \frac{h }{mL_x} \right) L_y\  ,
\label{incurrfi}
\eeq
where the expression in parenthesis is the {\it drift velocity} 
$v_d = h /{(mL_x)}$.
\section{Conclusions}

In this paper, we have considered effective field theories of the Calogero-Sutherland universality class. 
Their physical portrait is that of a $1D$ quantum compressible fluid, with non-linear waves of the
Benjamin-Ono type, involving both chiralities. This view stresses similarities and differences with the quantum
incompressible fluids that are familiar from the $QHE$ 
\cite{aw,bz}. These are $(c,{\overline c})=(1,1)$ $CFT$s of a compactified free boson field with extended 
symmetry $\winf \times {\overline \winf}$ and chiral and
antichiral sectors that are isomorphic. The Hilbert space and partition
function for these theories are known and their dynamics is given by the specific Calogero-Sutherland
hamiltonian. 

We have presented an extension of these theories to describe laminar quantum hydrodynamic
flows: a $2D$  effective field theory that describes charge transport along  
narrow systems as a direct product of two $1D$ ones, describing 
charge transport along the longitudinal dimension and confinement
in the transverse channel. The charge transport mechanism discussed here 
could be understood as the semiclassical flow of a planar laminar fluid 
with no classical viscosity, {\it i.e.},
with no friction among the quasi $1D$ layers of the system, taken as long strips
with very small width. Therefore, each strip travels at the same velocity as the
others and there is no vanishing of it at the boundaries of the sample
as a genuine classical Poiseuille flow would require. 
The vanishing of the electric current density at the boundaries is indeed a quantum 
mechanical property.

%
\def\NP{{\it Nucl. Phys.\ }}
\def\PRL{{\it Phys. Rev. Lett.\ }}
\def\PL{{\it Phys. Lett.\ }}
\def\PR{{\it Phys. Rev.\ }}
\def\IJMP{{\it Int. J. Mod. Phys.\ }}
\def\MPL{{\it Mod. Phys. Lett.\ }}

\end{document}